\documentclass{article}

\usepackage{times}
\usepackage{graphicx} 
\usepackage{subfigure} 

\usepackage{natbib}

\usepackage{algorithm}
\usepackage{algorithmic}
\usepackage{siunitx}
\usepackage{hyperref}

\usepackage{xcolor}
\usepackage{amsmath}
\usepackage{amsthm}
\usepackage{amsfonts}
\usepackage{amssymb}    
\usepackage{mathrsfs}
\usepackage{bbm, dsfont}
\usepackage[capitalise]{cleveref}
\usepackage{empheq}

\newcommand{\ie}[0]{\emph{i.e.},~}
\newcommand{\eg}[0]{\emph{e.g.},~}

\newcommand{\fix}[1]{}

\newcommand{\newLambda}[0]{\ensuremath{\Upsilon}}
\renewcommand{\Re}[0]{\ensuremath{\mathbb{R}}}
\newcommand{\xx}[0]{\ensuremath{{X}}}
\newcommand{\x}[0]{\ensuremath{{X}}}
\newcommand{\yy}[0]{\ensuremath{{Y}}}
\newcommand{\XX}[0]{\ensuremath{\newLambda}}
\newcommand{\nn}[1]{\ensuremath{^{(#1)}}}
\newcommand{\ff}[0]{\ensuremath{\mathsf{f}}}

\DeclareMathOperator*{\argmin}{arg\,min}

\newcommand{\norm}[1]{\lVert#1\rVert}
\newcommand{\Norm}[1]{\left\lVert#1\right\rVert}

\newcommand{\Idot}[2]{\left\langle #1 , #2 \right\rangle}
\newcommand{\idot}[2]{\langle #1 , #2 \rangle}

\newcommand{\iid}{\overset{iid}{\sim }}
\newcommand{\bw}{\ensuremath{\delta}}
\newcommand{\R}{\mathbb{R}}
\newcommand{\Z}{\mathbb{Z}}
\newcommand{\E}{\mathbb{E}}

\newcommand{\bZ}{{\bf{Z}}}

\newcommand{\calI}{\mathcal{I}}
\newcommand{\calX}{\mathcal{X}}

\newcommand{\calD}{\mathcal{D}}

\newcommand{\calN}{\mathcal{N}}

\newcommand{\vY}{\vec{Y}}

\newcommand{\tp}{\tilde{p}}
\newcommand{\tP}{\tilde{P}}

\newcommand{\hP}{\widehat{P}}

\newcommand{\hpsi}{\hat{\psi}}

\newcommand{\mpc}{~\ensuremath{\mathrm{h^{-1} Mpc}}~}
\newcommand{\avec}{\vec{a}}

\newcommand{\ud}{\mathrm{d}}

\newcommand{\where}{\mathrm{where }}

\newcommand{\Unif}{\text{Unif}}

\newcommand{\lp}[1]{}

\usepackage[accepted]{icml2016}

\icmltitlerunning{Cosmological Parameters from the Dark matter Distribution}

\begin{document} 

\twocolumn[
\icmltitle{Estimating Cosmological Parameters from the Dark Matter Distribution}

\icmlauthor{Siamak Ravanbakhsh$^{\star}$}{mravanba@cs.cmu.edu}
\icmlauthor{Junier Oliva$^{\star}$}{joliva@cs.cmu.edu}
\icmlauthor{Sebastien Fromenteau$^{\dagger}$}{sfroment@andrew.cmu.edu}
\icmlauthor{Layne C. Price$^{\dagger}$}{laynep@andrew.cmu.edu}
\icmlauthor{Shirley Ho$^{\dagger}$}{shirleyh@andrew.cmu.edu}
\icmlauthor{Jeff Schneider$^{\star}$}{jeff.schneider@cs.cmu.edu}
\icmlauthor{Barnab\'{a}s P\'{o}czos$^{\star}$}{bapoczos@cs.cmu.edu}
\icmladdress{$\star$ School of Computer Science,
Carnegie Mellon University, 5000 Forbes Ave., Pittsburgh, PA 15213, USA\\
$\dagger$ McWilliams Center for Cosmology, Department of Physics,
Carnegie Mellon University, Carnegie  5000 Forbes Ave., Pittsburgh, PA 15213, USA}

\icmlkeywords{machine learning, cosmology, deep learning, convolutional neural networks, volumetric data, distribution-to-real regression, dark-matter}

\vskip 0.3in
]

\begin{abstract} 
A grand challenge of the 21$^{st}$ century cosmology is to accurately estimate the cosmological parameters of our Universe.
A major approach in estimating the cosmological parameters is to use the large scale matter distribution of the Universe. Galaxy surveys provide the means to map out cosmic large-scale structure in three dimensions. Information about galaxy locations is typically summarized in a ``single'' function of scale, such as the galaxy correlation function or power-spectrum. 
We show that it is possible to estimate these cosmological parameters directly from the distribution of matter. This paper presents the application of deep 3D convolutional networks to volumetric
representation of dark-matter simulations as well as the results obtained using a 
recently proposed distribution regression framework, showing that machine learning techniques are comparable to, and can sometimes outperform, maximum-likelihood point estimates using ``cosmological models''.
This opens the way to estimating the parameters of our Universe with higher accuracy.
\end{abstract}

\section{Introduction}
\label{intro}
\begin{figure}[th!]
\centering
\includegraphics[width=.45\textwidth]{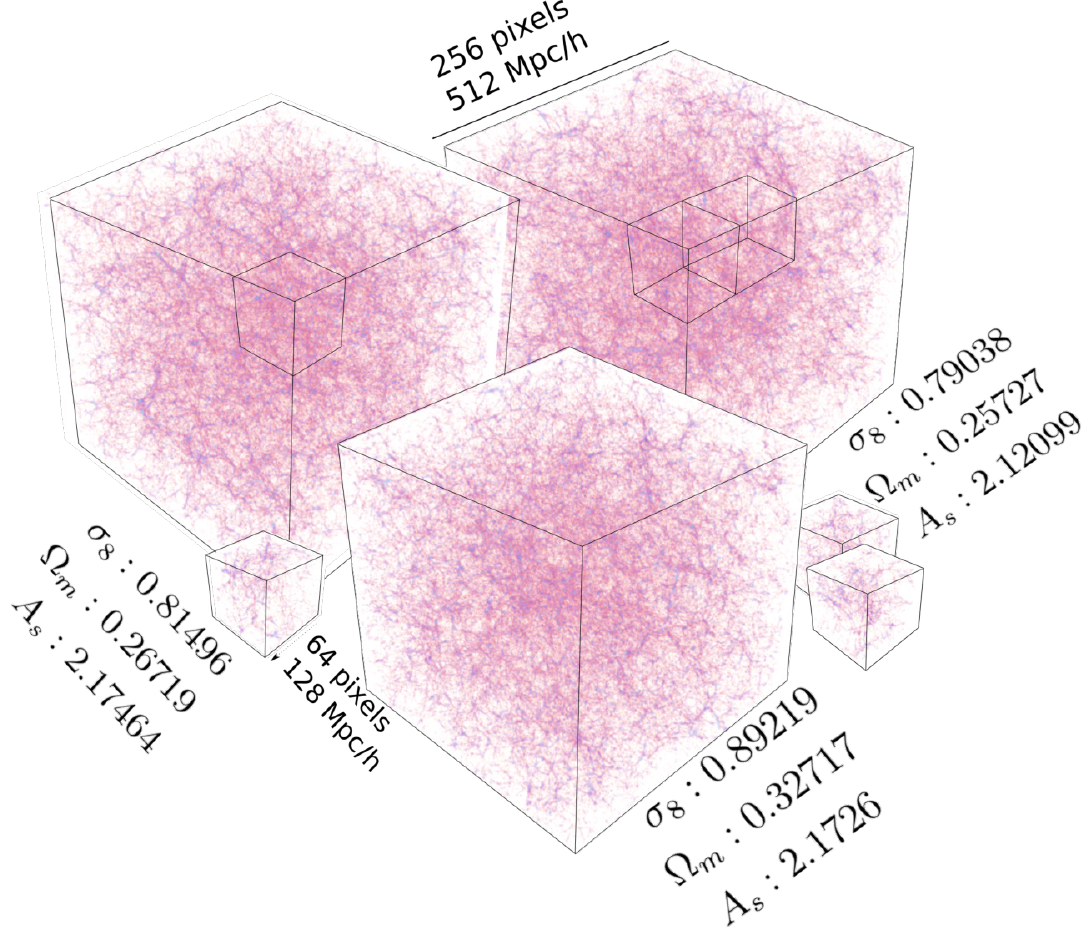}
\vspace{-1em}
\caption{\small{Dark matter distribution in three cubes produced using different sets of parameters. Each cube is divided into small sub-cubes for training and prediction.
Note that although cubes in this figure are produced using very different cosmological parameters in our constrained sampled set, the effect is not visually discernible.}}
\vspace{-1.5em}
\label{fig:cubes}
\end{figure}

The $21^\mathrm{st}$ century has brought us tools and methods to observe and analyze the Universe in far greater detail than before, allowing us to deeply probe the fundamental properties of cosmology.
We have a suite of cosmological observations that allow us to make serious inroads to the understanding of our own universe, including the cosmic microwave background (CMB) \cite{Planck_cosmo2015,Hinshaw:2012aka}, supernovae \cite{Perlmutter:1998np, Riess:1998cb} and the large scale structure of galaxies and galaxy clusters \cite{2df, sdss, wigglez}. 
In particular, large scale structure involves measuring the positions and other properties of bright sources in great volumes of the sky. The amount of information is overwhelming, and modern methods in machine learning and statistics can play an increasingly important role in modern cosmology. For example, the common method to compare large scale structure observation and theory is to compare the compressed two-point correlation function of the observation with the theoretical prediction (which is only correct up to a certain physical separation scale). We argue here that there may be a better way to make this comparison. 


The best model of the Universe is currently described by less than 10 parameters in the standard $\Lambda$CDM cosmology model, where CDM stands for cold dark matter and $\Lambda$ stands for the cosmological constant. The $\Lambda$CDM parameters that are important for this analysis include the matter density $\Omega_m \approx 0.3$ (normal matter and dark matter together constitute $\sim 30\%$ of the energy content of the Universe), the dark energy density $\Omega_\lambda \approx 0.7$ ($\sim$70\% of the energy content of the Universe is a dark energy substance that pushes the content of the universe apart), the variance in the matter over densities $\sigma_8 \approx 0.8$ (measured on the matter power spectrum smoothed over 8 \mpc spheres), and the current Hubble parameter $H_0 =100 h \approx 70 \mathrm{km/s/Mpc}$ (which describes the present rate of expansion of the Universe).  $\Lambda$CDM also assumes a flat geometry for the Universe, which requires $\Omega_\Lambda = 1 - \Omega_m$ ~\citep{dodelson2003modern}. Note that the unit of distance megaparsec/$h$ (\mpc) used above is time-dependent, where $1 \mathrm{Mpc}$ is equivalent to $3.26\times 10^6$ light years and 
$h$ is the dimensionless Hubble parameter that accounts for the expansion of the universe.

The expansion of the Universe stretches the wavelength, or \emph{redshifts}, the light that is emitted from distant galaxies, with the amount of change in wavelength depending on their distances and the cosmological parameters.  Consequently, for a fixed cosmology we can use the directly observed redshift $z$ of galaxies as a proxy for their distance away from us and/or the time at which the light was emitted.  

Here, we present a first attempt at using advanced machine learning to predict cosmological parameters directly from the distribution of matter. The final goal is to apply such models to 
produce better estimates for cosmological parameters in our Universe. 
In the following, \Cref{sec:results} presents our main results. 
\Cref{sec:methods} elaborates on the simulation and cosmological analysis procedures as well as machine learning techniques used to obtain these estimates.   

\section{Results}\label{sec:results}
To build the computational model, we rely on direct dark matter simulations produced using different cosmological parameters and random seeds. 
We sample these parameters within a very narrow range that reflects the uncertainty of our current best estimates of these parameters for our universe from real data, in particular the \citet{Planck_cosmo2015} CMB observations. Our objective is to show that 
it is possible to further improve the estimates in this range, for simulated data, using a deep convolutional neural network (conv-net). 

We consider two sets of simulations: the \textbf{first set} contains only one snapshot of the dark matter distribution at the present day. The following cosmological parameters are varied across simulations: I) mass density $\Omega_m$; II) $\sigma_8$
(or alternatively, the amplitude of the primordial power spectrum, $A_s$, which can be used to predict $\sigma_8$).

\begin{figure}
\centering
\includegraphics[width=.5\textwidth]{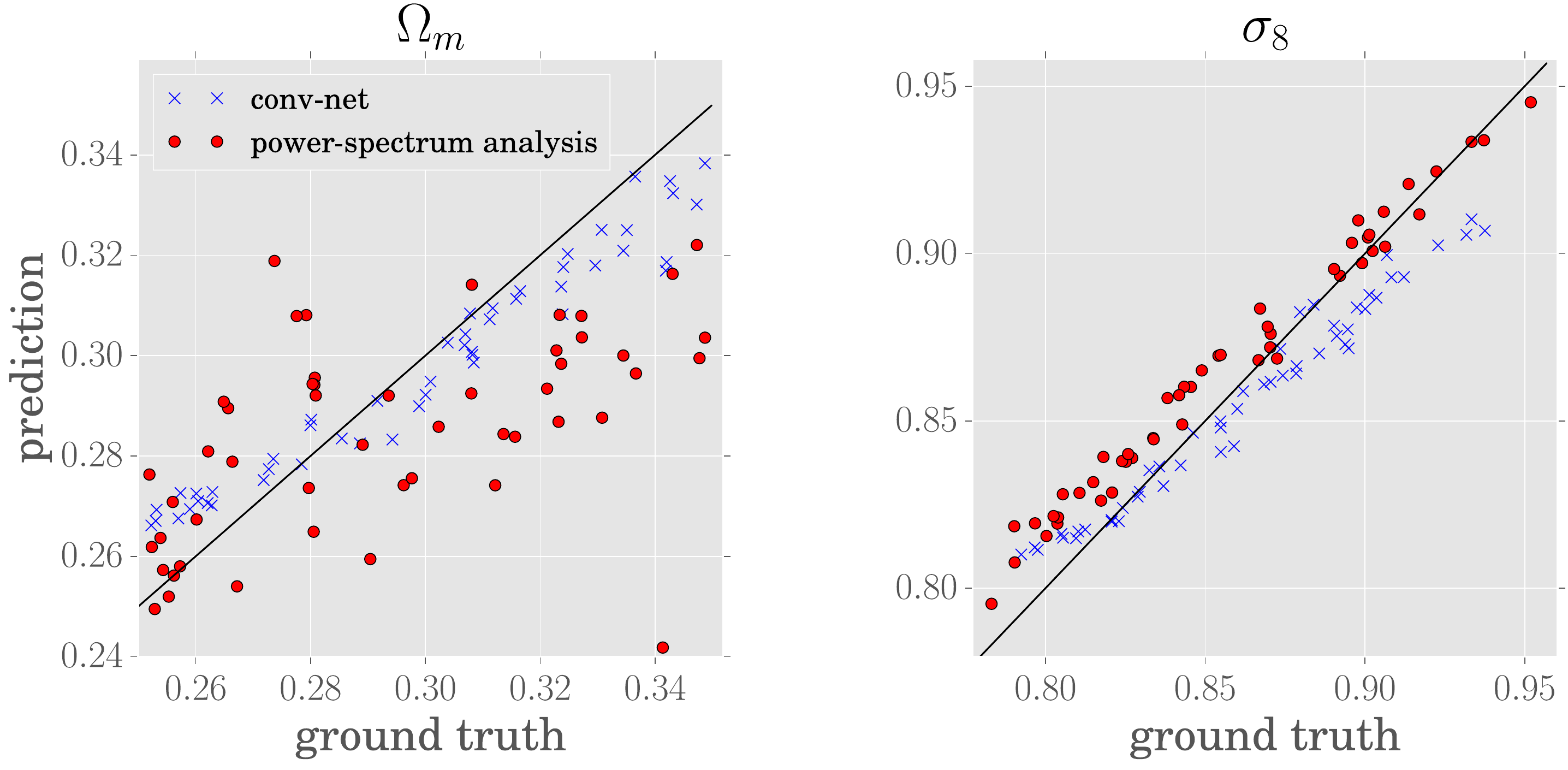} 
\vspace{-1.5em}
\caption{\small{
Prediction and ground truth of $\Omega_m$ and $\sigma_8$ using 3D conv-net and analysis of the power-spectrum on 50 test cube instances. 
}
\vspace{-1.5em}
}\label{fig:pred}
\end{figure}

Here, each training and test instance is the output of an N-body simulation
with millions of particles in a box or ``cube'' that is tens of \mpc across.
All the simulations in this dataset are recorded at the present day -- \ie redshift $z = 0$.
\Cref{fig:cubes} shows three cubes with their corresponding cosmological parameters. As is evident from this figure, distinguishing the constants using 
visual clues is challenging. Importantly, there is substantial variation among cubes even with similar cosmological parameters, since the initial conditions are chosen randomly in each simulation. 
In all experiments, we use $90\%$ of the data for training and the remaining $10\%$ for testing.

We compare the performance of the conv-net to a standard cosmology analysis based on the standard maximum likelihood fit to the matter power spectrum~\citep{dodelson2003modern}. \Cref{fig:pred} presents our main result, the prediction versus the ground truth for the cosmological parameters using both methods.
We find that the maximum likelihood prediction for $(\sigma_8, \Omega_m)$ has an average relative error of $(0.013,0.072)$, respectively.\footnote{Relative error for ground truth $\Omega_m$ and the prediction $\widehat{\Omega}_m$ are defined as $\left(|\Omega_m - \widehat{\Omega}_m|\right)/\Omega_m$.} In comparison, the conv-net has an average relative error of $(0.012, 0.028)$, which has a clear advantage in predicting $\Omega_m$.  Predictions for conv-net are the mean-value of the predictions on smaller 128 \mpc sub-cubes.
On these sub-cubes, the conv-net has a relatively small standard deviation of $(0.0044, 0.0032)$, indicating only small variations in predictions using much smaller sub-cubes.  We have not performed a maximum likelihood estimate on these small sub-cubes, since the quality of the results would be drastically limited by sample variance.\footnote{For the power spectrum analysis there is a strong degeneracy in the $(\sigma_8, \Omega_m)$ plane on small scales: larger (smaller) values of $\sigma_8$ combined with smaller (larger) $\Omega_m$ predict comparable power spectra.  This provides a small bias to the maximum likelihood estimate.}
We also observed that changing the size of these sub-cubes by a factor of two did not significantly affect conv-net's prediction accuracy;
see the Appendix A for details.

\begin{figure}[h]
\centering
\hbox{\includegraphics[width=.4\textwidth]{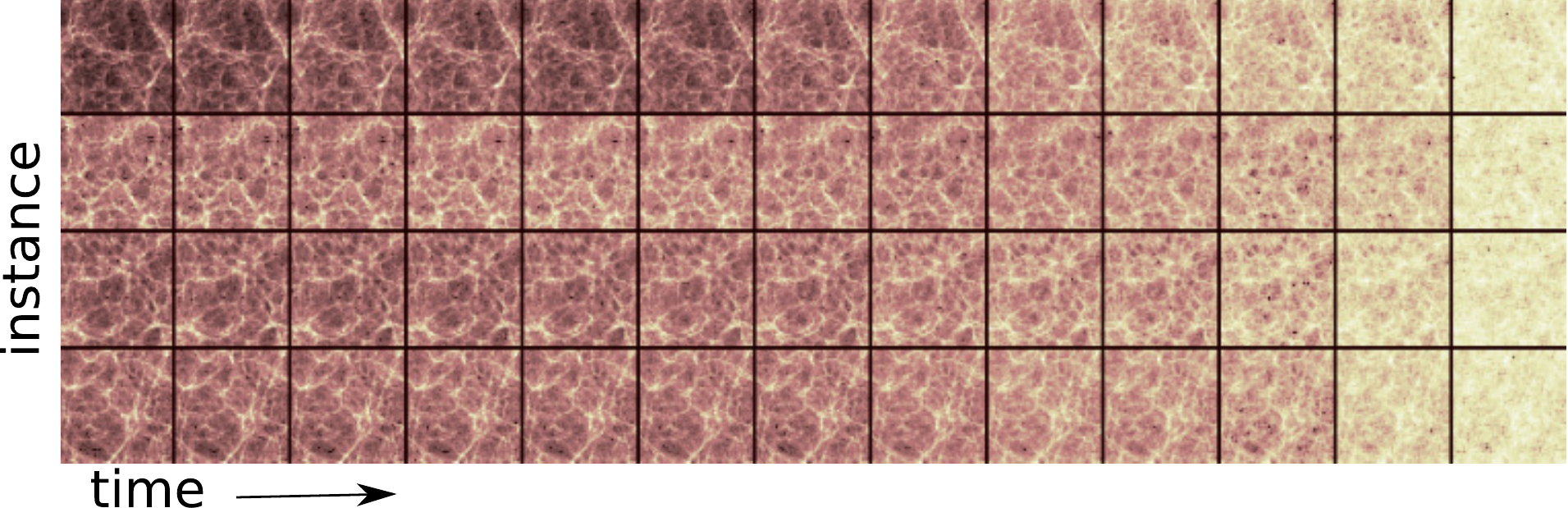}}
\vspace{-1em}
\caption{\small{Log-density of dark matter at different redshifts. Each row shows a slice of 
a different 3D cube. From left to right the redshift increases in $\mathrm{1 Gyr}$ steps.}
\vspace{-1.5em}
}\label{fig:redshift_data}
\end{figure}

\begin{figure}
\centering
\includegraphics[width=.4\textwidth]{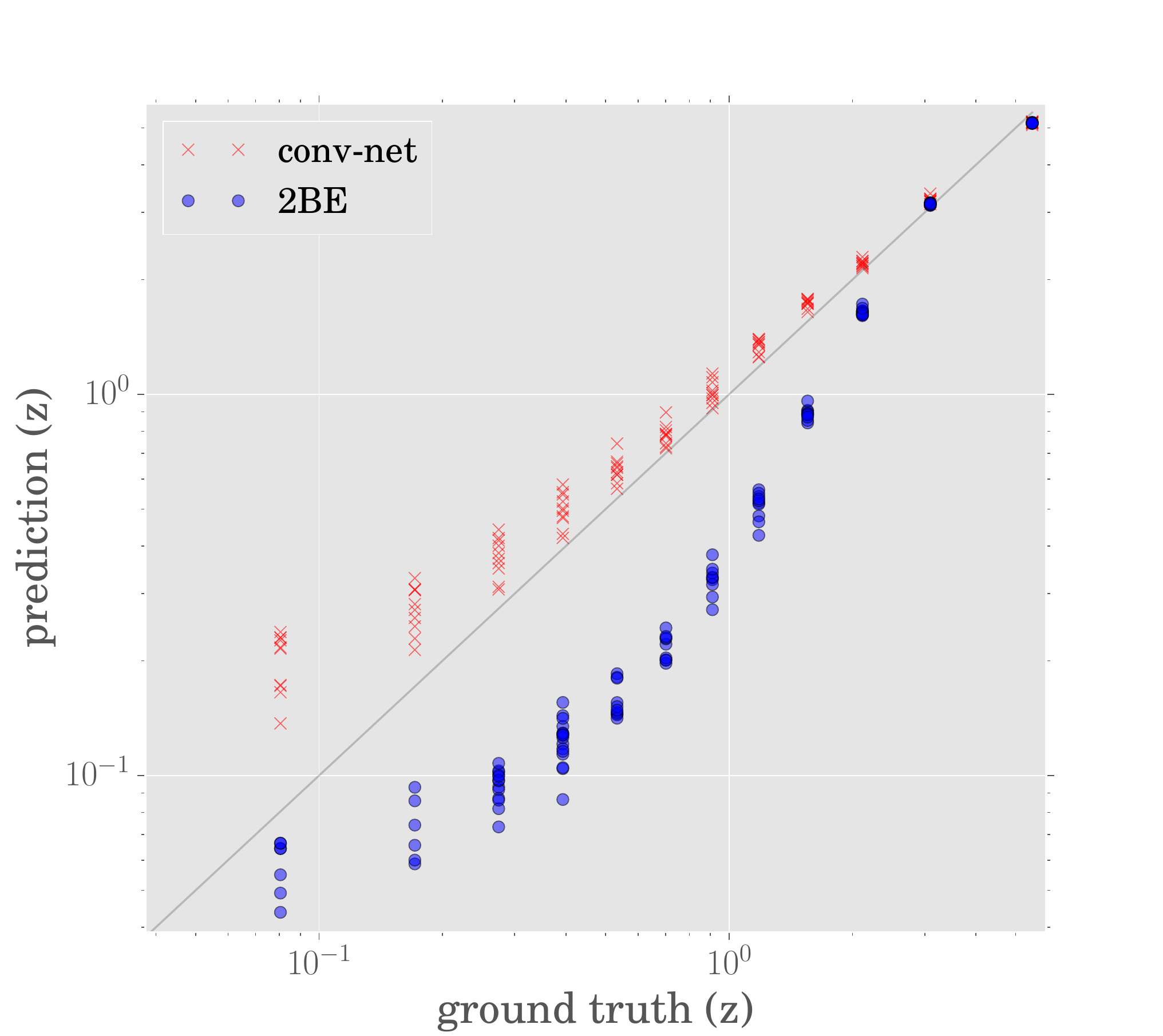} 
\caption{\small{
Prediction and ground truth of redshift $z$ on test instances for both 3D conv-net and double-basis estimator (2BE).
}}\label{fig:pred_redshift}
\vspace{-1.5em}
\end{figure}

The \textbf{second dataset} contains 100 simulations using a more sophisticated simulation code~\citep{Trac2015}, where each simulation is recorded at 13 different redshifts $z \in [0,6]$; see \Cref{fig:redshift_data}. Simulations in this set use fewer particles and
since the distribution of matter at different redshifts is substantially different (compared to the effect of cosmological parameters in the first dataset) we are able to produce reasonable estimates of the redshift using the distribution-to-real framework of~\citet{oliva2014fast} as well as a 3D conv-net.
\Cref{fig:pred_redshift} reports both results for the training and test sets.

\section{Methods}\label{sec:methods}
We review the procedure for dark matter simulations in \Cref{sec:simulations} and outline the standard cosmological likelihood analysis in \Cref{sec:twopoint}.
\Cref{sec:inv} and \Cref{sec:conv-net} detail our deep conv-net applied to the data and our approach to predicting the redshift using a double-basis estimator.
\Cref{sec:redshift} describes the details of the redshift estimation.

\subsection{Simulations}\label{sec:simulations}

Simulations play an important part in modern cosmology studies, particularly in order to model the non-linear effects of general relativity and gravity, which are impossible to take into account in a simpler analytic solution. Consequently, significant effort has been made in the last few decades to obtain a large number of realistic simulations as a function of the cosmology parameters. The simulations also provide a useful test for supervised machine learning techniques. 

In order to apply a Machine Learning process in cosmological parameter estimation we need to generate a huge amount of simulations for the training set. 
Moreover, it is important to generate big volume simulation boxes in order to accurately reproduce the statistics of large scale structures.
There are several algorithms for calculating the gravitational acceleration in N-body simulations, ranging from slow-and-accurate to fast-and-approximate. The equations of motion for the N particles are solved in discrete time steps to track the nonlinear trajectories of the particles.

\begin{figure}[t]
\centering
\includegraphics[width=.45\textwidth]{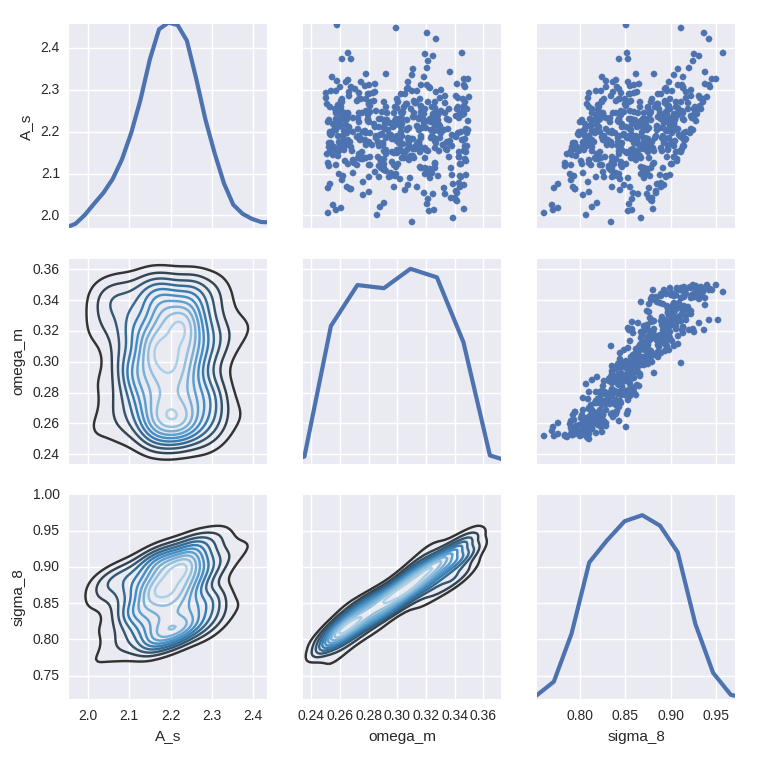}
\vspace{-1em}
\caption{\small{Distribution of cosmological parameters in the first set of simulations.}}
\vspace{-1.5em}
\label{fig:dist}
\end{figure}


As we are interested in large scale statistics, for the \textbf{first dataset} we use the COmoving Lagrangian Acceleration (COLA) code \citep{Tassev2013,Koda2015}.  
The COLA code is a mixture of N-body simulation and second order Lagrangian perturbation theory.
This method conserves the N-body accuracy at large scale and agrees with the non-linear power spectrum (see \Cref{sec:twopoint}) that can be obtained with ultra high-resolution pure N-body simulations~\citep{Springel2005} at better than $95\% $ up to $k \sim 0.5 h \mathrm{Mpc^{-1}}$.

For the first study we generate 500 cubic simulations with a size of $512 \mpc$ with $512^3$ dark matter particles, evolving the simulation until redshift $z=0$. The mass of these particles varies with the value of $\Omega_m$ from $m_p\sim 6.5 \times 10^{10}$ to $m_p\sim9.5 \times 10^{10}\mathrm{h^{-1}M_{\odot}}$, where $M_{\odot}$ is a solar mass. 
We start the simulations at a redshift of $z\sim 20$ and use 20 steps up to the final redshift $z=0$.\footnote{Corresponding to a scale factor of $a=0.05$, as advocated in \citet{Izard2015}.} Each box is generated using a different seed for the random initial conditions.\footnote{ This random seed is generated by an adjusted version of the \textsc{2LPTic} code \cite{Koda2015}.} The Hubble parameter used for all simulations is $H_0 = 70 \mathrm{km / s /Mpc}$.\footnote{We use a scalar perturbation spectral index of $0.96$ for all the simulations and a cosmological constant with $\Omega_{\Lambda}=1-\Omega_{m}$ in order to conserve a flat universe. Each simulation on average requires 6 CPU hours on 2GHz processors and the final raw snapshot is about 1GB in size.}

Motivated by the \textsc{Planck} results~\cite{Planck_cosmo2015}, we use a Gaussian distribution for the amplitude of the initial scalar perturbations $\ln(10^{10} A_s )=3.089\pm 0.036$ and a flat distribution in the range $\left[ 0.25;0.35 \right]$ for $\Omega_{m}$.
Note that \textsc{Planck} results arguably give us the best constraints on the parameters of our 
Universe, limiting our simulations mostly to uncertain regions of the parameter space.
The value for $\sigma_8$ is obtained by calculating the convolution of the linear power spectrum with a top hat window function with a radius of $8 \mpc$, using
the \textsc{Camb} code; see \Cref{sec:twopoint} for power-spectrum.
\Cref{fig:dist} shows the distribution of the three parameters (two independent, one derived) that are 
varying across simulations.   


The simulations in the \textbf{second dataset} are based on a particle-particle-particle-mesh ($\mathrm{P^3M}$) algorithm from \citet{Trac2015}.\footnote{The long-range potential is computed using a particle-mesh algorithm where Poisson's equation is efficiently solved using Fast Fourier Transforms. The short-range force is computed for particle-particle interactions using direct summation.} Each simulation is computed in 13 time steps of 1 gigayear ($\mathrm{Gyr}$), using $128^3$ particles in boxes with sizes of 128 \mpc using the standard $\Lambda CDM$ cosmology. 

\subsection{Two-Point Correlation and Maximum Likelihood Power Spectrum Analysis}
\label{sec:twopoint}
A commonly used measurement for analysis of the distribution of matter
is the two-point correlation function $\xi(\vec{r})$, measuring 
the excess probability, relative to a random distribution, of finding two points in the matter distribution at the volume elements $dV_1$ and $dV_2$ separated by a vector distance $\vec{r}$ -- that is we have 
\begin{equation}
dP_{12}(\vec{r}) = n^2\left(1+\xi(\vec{r})\right)dV_1dV_2,
\end{equation}
where $n$ is the mean density (number of particles divided by the volume), and
$n^2 dV_1dV_2$ in the equation above measures the probability of finding two points 
 in $dV_1$ and $dV_2$ at vector distance $\vec{r}$.
Under the cosmological principle the Universe is statistically isotropic and homogeneous, therefore the correlation function only depends on the distance $r=|\vec{r}|$.
The matter power spectrum $P_m(k)$ is the Fourier transform of the correlation function, where $k = |\bf k|$ is the magnitude of the Fourier basis vector.

The form of the power spectrum as a function of $k$ depends on the cosmological parameters, in particular $\sigma_8$ and $\Omega_m$. For a larger (smaller) $\sigma_8$ the amplitude of the power spectrum smoothed on the scale of $8 \mpc$ increases (decreases).  Similarly, larger $\Omega_m$ shifts power into smaller scales. 

Given the output of an N-body simulation at $z=0$, we evaluate the ``empirical'' power spectrum $\hat P(k)$ of the dark matter distribution.%
\footnote{
Given the $\Lambda$CDM cosmology model, there is a constraint in the parameter space $(A_s, \sigma_8, \Omega_m)$, which we utilize to only require fitting to the parameters $(\sigma_8, \Omega_m)$ -- \ie treating $A_s$ as a deterministic derivative.
}
For a set of cosmological parameters $\yy = (\sigma_8, \Omega_m)$ we can obtain 
the predicted (theoretical) matter power spectra $P_m(k, \yy)$.\footnote{We use the linear Boltzmann code \textsc{Camb}~\citep{Lewis1999}, supplemented with the empirically calibrated non-linear corrections obtained from \textsc{Halofit} \citep{Smith:2002dz}. This is basically an accurate estimate of the \textit{average} power spectra, \textit{if} our training 
dataset contained many simulations with the same cosmological parameters and different initial conditions.} 
This theoretical average is produced using our ``physical model'', rather than
the training data. After obtaining an estimate of the covariance using additional training simulations, for each test cube, we can find the parameter $\yy$ 
that maximizes its Gaussian likelihood.

To define this Gaussian likelihood of the empirical power spectra based on its theoretical value $\mathcal L(\hat P_m(k) | P_m(k, \yy) )$, we discretize the power spectrum to equally spaced bins in $\log k$.  We the estimate the covariance matrix of this Gaussian using 20 different simulations with the fixed cosmology of $(\sigma_8, \Omega_m) = (0.812, 0.273)$. Note that each of these is using different random initial conditions to obtain an estimate of the sample variance.\footnote{
These particular parameters provide the best-fit Lambda-CDM values to the data from the Planck satellite telescope, which is the state-of-the-art measurement of the cosmic microwave background.}
The sample variance on scales of $k\lesssim 0.1 \mathrm{Mpc}$ gives a large uncertainty in the estimate of $\hat P_m(k)$ at scales $\gtrsim 100 \mpc$ in real-space, which corresponds to approximately $20\%$ of the entire simulation box. This limits the inferences we can draw from large scales in the dark matter simulation.

We then maximize the likelihood function over $\yy$ using the downhill simplex method~\citep{nelder1965simplex} to obtain an estimate $\hat \yy$ that can be compared to the ground truth cosmological parameter values that are known from the simulations. \footnote{While this differs from common cosmological analyses that calculate the posterior probability distribution $P(\yy | D)$ using Bayesian techniques via software such as \textsc{CosmoMC}~\citep{Lewis:2002ah}, it gives a reasonable point estimate of the parameters that can be compared to the results of the conv-nets.}

\subsection{Invariances of the Distribution of Matter}
\label{sec:inv}
Modern cosmology is built on the cosmological principle that states at large scales, the distribution of matter
in the Universe is homogeneous and isotropic~\citep{ryden2003introduction}, which implies shift, rotation and reflection invariance of the distribution of matter. These invariances have also made the two-point correlation function --as a shift, rotation and reflection invariant measurement-- an indispensable tool in cosmological data analysis. Here, we intend to go beyond this measure.   
Let $\xx$ denote a cube and $\yy = (\Omega_m, \sigma_8)$ the corresponding dependent variable. 
The existence of invariance in the data means $p(\yy \mid \xx) = p(\yy \mid \mathrm{transform}(\xx))$, where the invariance identifies the valid transformations. 

In machine learning, and in particular deep learning, several recent works have attempted to identify and model the data invariances and its symmetries~\citep[\eg][]{gens2014deep,cohen2014learning}. However, due to inefficiency of current techniques, any known symmetry beyond translation invariance is often enforced by data-augmentation~\citep[\eg][]{krizhevsky2012imagenet}; see \citep{ dieleman2015rotation} for an application in astronomy. 
Data-augmentation is the process of replicating data by invariant transformations.

In the original representation of cubes, particles are fully interchangable and a source of redundancy is due to this permutation invariance.
For conv-nets, prior to data augmentation, we transform this data to volumetric form, where a 3D histogram of $d^3$ voxels represents the normalized density of the matter for each cube. For the first and second datasets this resolution (in proportion to the number of particles and the size of these cubes) is set to $d=256$ and $d=64$ respectively, which means each voxel is $2 \mpc$ along each edge.
A normalization step ensures that the model generalizes to simulations with different number of particles as long as densities remain non-degenerate.
In the first dataset we further break down each of the $500$ simulation cubes to $64^3$-voxel sub-cubes, corresponding to $128^3 (\mpc)^3$. This is in order to obtain more training instances for our conv-net; see \Cref{fig:cubes}

Translation invariance is addressed by shift-invariance of the convolutional parameters.
We augment both datasets with symmetries of a cube. This symmetry group has 48 elements:
 6 different \SI{90}{\degree} rotations and $2^3 = 8$ different axis-reflections of each sub-cube.

The combination of data-augmentation and using ``sub''-cubes increases the training data $\mathcal{S} = \{(\xx\nn{1}, \yy\nn{1}),\ldots,(\xx\nn{N}, \yy\nn{N})\}$ to have $N > 10^6$ and $N > 62000$ instances for the first and second dataset respectively, where in the following we use $\xx \in \XX = \Re^{64^3}$ to denote a (sub-)cube from either dataset. 
To see if the data-augmentation has indeed produced the desirable invariance, we
predicted both $\Omega_m$ and $\sigma_8$ using 48 replicates of each sub-cube. The average standard deviation in these predictions is .0013 and .0017 respectively, \ie small
compared to .029 and .039, their respective standard deviations over the whole test-set.



\subsection{Deep Convolutional Network for Volumetric Data}\label{sec:conv-net}
Our goal is to learn the model parameters $\theta^* \in \Theta$ for an expressive class of functions $\ff_{\theta}: \XX \to \Re^2$, so as to minimize the expected loss 
$\mathbb{E}_{\xx, \yy} [\ell(\ff(\xx) - \yy)]$ where $\ell(\Re^2) \to \Re$ is a loss function --- \eg we use the L1 norm. 
However, due to the unavailability of $p(\xx, \yy)$, 
a common practice is to minimize the empirical loss
$\sum_{(\xx\nn{n}, \yy\nn{n}) \in \mathcal{S}} \ell(\ff(\xx\nn{n}) - \yy\nn{n})$ with an eye towards
generalization to new data, which is often enforced by regularization.

Our function class is the class of a deep convolutional neural network \citep{lecun2015deep, bengio2009learning}.
Conv-nets have been mostly applied to 2D image data in the past. Beside applications in video processing --with two image dimensions and time as the third dimension-- application of conv-nets to volumetric data are very recent and mostly limited to 3D medical image segmentation~\citep[\eg][]{kamnitsas2015multi,roth2015deep}. 

\begin{figure}[h]
\centering
\includegraphics[width=.45\textwidth]{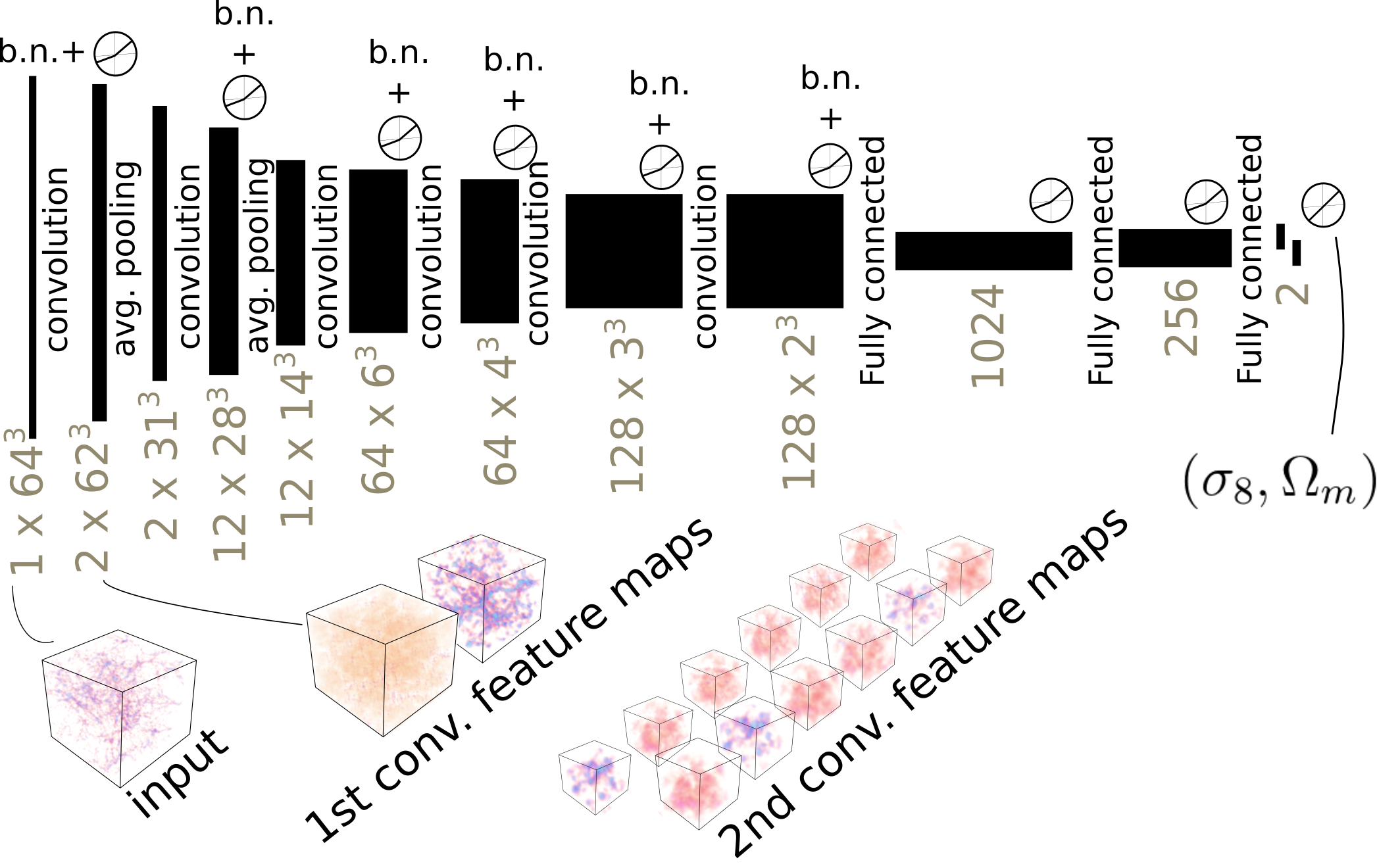}
\caption{\small{The architecture of our 3D conv-net.
The model has six convolutional and 3 fully connected layers. The first two
convolutional layers are followed by average pooling. All layers, except the final layer, use leaky rectified linear units,
and all the convolutional layers use batch-normalization (b.n.).}
}\label{fig:arc}
\vspace{-1.5em}
\end{figure}

\Cref{fig:arc} shows the architecture of our model. 
A major restriction when moving from 2D images to volumetric data is the substantial increase in the size of the input, which in turn restricts the number of feature-maps at the first layers of the conv-net. This memory usage is further amplified by the fact that in 3D convolution the advantage of using FFT is considerable. However, FFT-based convolution requires larger memory compared to its time domain counterpart. 

In designing our network we identified several choices that are critical in obtaining the results reported in \Cref{sec:results}:
\noindent 
I) We use \textit{\textit{Leaky} rectified linear unit (ReLU).}~\citep{maas2013rectifier}. This significantly speeds up the learning compared to non-leaky variation. We used the leak parameter $c = .01$ in $\ff(\x) = \max(0,\x) - c$. 

\noindent
II) We used \textit{Average pooling} in our model and could not learn a meaningful model using max-pooling (which is often used for image processing tasks). One explanation is that with the combination of ReLU and average pooling, activity at higher layers of the conv-net signifies the \textit{weighted sum} of the dark-matter mass at particular regions of the cube. This information (total mass in a region) is lost when using max-pooling. Here, both pooling layers are sub-sampling by a factor of two along each dimension.

\noindent
III) \textit{Batch normalization}~\citep{ioffe2015batch} is necessary to undo the internal covariate shift and stabilize the gradient calculations. 
The basic idea is to normalize the output of each layer --with an online estimate of mean and variance for all the training data at that layer-- before applying the non-linearity.
\footnote{Without using batch-normalization, we observed shooting gradients early during the training.Batch-normalization would not be critical in a more shallow network. However, we observe consistent --although sometimes marginal-- improvement   by increasing the number of layers in our conv-net up to its current value.
In using batch-normalization, we normalize the values across all the voxels of each feature-map. However, since due to memory constraints the number of training instances in each mini-batch is limited, batch-normalization across the fully connected layers introduces relatively large oscillations during learning. For this reason, we limit the batch-normalization to convolutional layers.}

Regularization is enforced by ``drop-out'' at fully connected layers, where 
$50\%$ of units are ignored during each activation, in order to 
reduce overfitting by preventing co-adaptation~\citep{hinton2012improving}.
For training with backpropagation, we use
Adam~\citep{kingma2014adam} with a learning rate of $.0005$ and
first and second moment exponential decay rate of $.9$
and $.999$, respectively.

\begin{figure}[h!]
\centering
\hbox{\includegraphics[width=.5\textwidth]{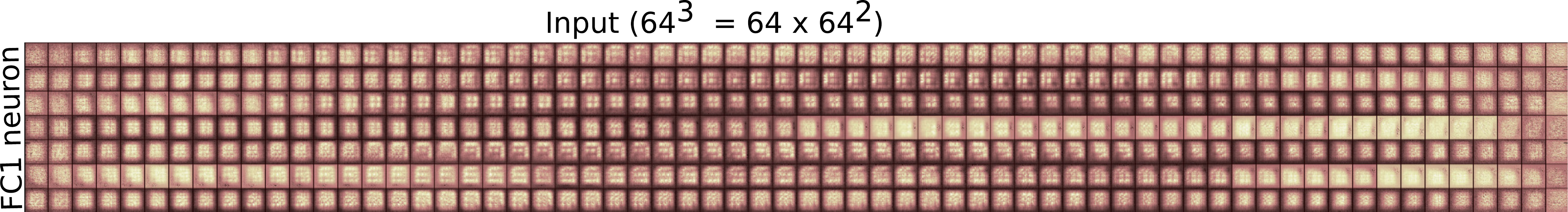}}
\hbox{\includegraphics[width=.5\textwidth]{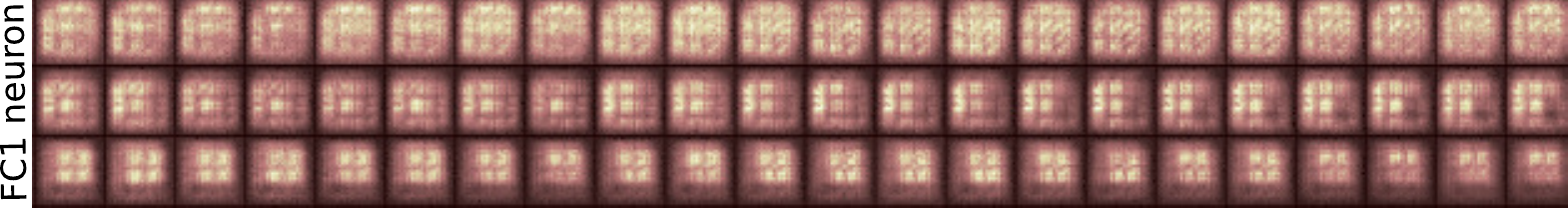}}
\caption{\small{ (top) visualization of inputs that maximize the activation of 7/1024 units (corresponding to seven rows) at the first fully connected layer.
In this figure, we have unwrapped the maximizing input sub-cubes for better visualization. 
(bottom) magnified portion of the top row.
}
}\label{fig:rep}
\end{figure}

\subsubsection{Visualization}
A common approach to visualizing the representation
learned by a deep neural network is to maximize the activation of a particular 
unit while treating the input $\xx$ as the optimization variable \citep{erhan2009visualizing,simonyan2013deep}
$$
\xx^* = \arg\max_{\xx} \;s.t.  \;\x_{l,i} \quad \|\xx\|_2 \leq \zeta
$$
where $\x_{l,i}$ is the $i^{th}$ unit at layer $l$ of the conv-net and $\zeta >0$ is a constant.
\Cref{fig:rep} shows the representation learned by seven units at the first fully
connected layer of our model.\footnote{Since the input to our conv-net is a distribution it seems more appropriate to bound $\xx$ by $\|\xx\|_1 = 1$ and $\x_{i} > 0\; \forall i$. However, using penalty method for this optimization did not produce visually meaningful features.} The visualization suggests that the conv-net has learned to identify various patterns involving periodic concentration of mass as a key feature in predicting $\Omega_m$ and $\sigma_8$.


\subsection{Estimating the Redshift}
\label{sec:redshift}
We applied the conv-net of the previous section to estimate the redshift in our second dataset.
Since this is an easier task, we removed two fully connected layers, without losing prediction power. All the other settings in training are kept the same.
For this dataset we could also obtain good results using the Double-Basis Estimator, described in the following section.

\subsubsection{Distribution to Real Regression}
We analyzed the use of distribution-to-real regression \cite{poczos2013distribution} and the Double-Basis Estimator (2BE) \cite{oliva2014fast} for predicting cosmological parameters. Here, we take sub-cubes of simulation snapshots to be sample sets from an underlying distribution, and regress a mapping that maps the underlying distribution to a real-value (in this case the redshift of the simulation snapshot). In other words, we consider our data to be $\calD = \{(\calX_i,Y_i)\}_{i=1}^N$, where $\calX_i = \{ X_{ij}\in \R^3 \}_{j=1}^{n_i} \iid P_i$. We look to estimate a mapping $Y_i = f(P_i) + \epsilon_i$, where $\epsilon_i$ is a noise term \cite{oliva2014fast}. 

Roughly speaking, the 2BE operates in an approximate primal space that allows one to use a kernelized estimator on distributions without computing a Gram matrix. The 2BE uses: 

\noindent
I) An orthonormal basis so that we can estimate the $L_2$ distance on two distributions, $\norm{P_i-P_j}_2$, as the Euclidean distance of finite vectors of their projection coefficients onto a finite subset of the orthonormal basis, $\norm{\avec(P_i)-\avec(P_j)}$.

\noindent 
II) A random basis to approximate kernel evaluations on distributions $K(P_i,P_j)$ as the dot product of finite vectors of random features on the respective projection coefficients of the distributions, $z(\avec(P_i))^Tz(\avec(P_j))$.

Using these two bases, the 2BE is able to regress a nonparametric mapping efficiently. In short, the 2BE estimates a real valued response, $Y_i$, as $Y_i \approx \psi^T z(\avec(P_i))$, where $z(\avec(P_i))$ are the aforementioned random features of projection coefficients, and $\psi$ is a vector of model parameters that are optimized over. We expound on the details below.

\noindent \textbf{Orthonormal Basis.}
We use orthonormal basis projection estimators \cite{tsybakov2008introduction} for estimating densities of $P_i$ from a sample $\calX_i$. Let $\newLambda = [a,b]$ and suppose that $\newLambda^l \subseteq \R^l$ is the domain of input densities. If $\{\varphi_i\}_{i\in\Z}$ is an orthonormal basis for $L_2(\newLambda)$, then the tensor product of $\{\varphi_i\}_{i\in\Z}$ serves as an orthonormal basis for $L_2(\newLambda^l)$; that is,
\begin{gather}
\{\varphi_\alpha\}_{\alpha\in\Z^l} \; \mathrm{where} \; \varphi_\alpha(x) = \prod_{i=1}^l \varphi_{\alpha_i}(x_i),\ x\in \newLambda^l \label{eq:basisprod}
\end{gather}
serves as an orthonormal basis (so we have $\forall \alpha,\rho\in \Z^l,\ \langle\varphi_\alpha,\varphi_\rho \rangle= I{\{\alpha=\rho\}}$).

Let $P\in \calI \subseteq L_2(\newLambda^l)$, then
\begin{gather}
p(x)=\sum_{\alpha\in\Z^l} a_\alpha(P)\varphi_\alpha(x)  \ \where\\
\quad a_\alpha(P) = \langle\varphi_\alpha, p\rangle = \int_{\newLambda^l} \varphi_\alpha(z)\ud P(z)\ \in \R \nonumber.
\end{gather}
Here, $p(x)$ denotes the probability density function of the distribution $P$.
If the space of input densities, $\calI$, is in a Sobolov ellipsoid type space; see \cite{ingster2011estimation,laurent1996efficient,oliva2014fast} for details. We can effectively approximate input densities using a finite set of empirically estimated projection coefficients.
Given a sample $\calX_i = \{X_{i1},\ldots,X_{in_i}\}$ where $X_{ij}\iid P_i \in \calI$, let $\hP_i$ be the empirical distribution of $\calX_i$; i.e. $\hP_i(X=X_{ij})=\frac{1}{n_i}$. Our estimator for $p_i$ will be:
\begin{align}
\tp_i(x) = \sum_{\alpha\in M}a_\alpha(\hP_i)\varphi_\alpha(x) \quad \where \label{eq:coef-est}\\
a_\alpha(\hP_i) = \int_{\newLambda^l} \varphi_\alpha(z)\ud \hP_i(z) = \frac{1}{n_i}\sum_{j=1}^{n_i} \varphi_\alpha(X_{ij}) \label{eq:coef-hat}.
\end{align}
Choosing $M$ optimally can be shown to lead to $\E[\norm{\tp_i-p_i}_2^2]=O(n_i^{-\frac{2}{2+\gamma^{-1}}})$, where $\gamma^{-1}$ is a smoothing constant \cite{nussbaum1983optimal}.

\noindent \textbf{Random Basis.}
Next, we use random basis functions from Random Kitchen Sinks (RKS) \cite{rahimi2007random} to compute our estimate of the response. In particular, we consider the RBF kernel 
\begin{align*}
K_\bw(x,y) = \exp\left(-\frac{\norm{x-y}^2}{2\bw^2}\right)
\end{align*}
where $x,y \in \R^d$ and $\bw\in\R$ is a bandwidth parameter. 
\citet{rahimi2007random} shows that for a shift-invariant kernel, such as $K_\bw$:
\begin{align}
&K_\bw(x,y) \approx z(x)^Tz(y),\ \where\\
&z(x) \equiv \sqrt{\tfrac{2}{D}}\left[\cos(\omega_1^Tx+b_1) \cdots \cos(\omega_D^Tx+b_D)\right]^T \label{eq:rksrfeat}
\end{align}
with $\omega_i \stackrel{iid}{\sim} \calN(0,\bw^{-2}I_d)$, $b_i \stackrel{iid}{\sim} \Unif[0,2\pi]$. 
The quality of the approximation will depend on the number of random features $D$ as well as other factors, see \cite{rahimi2007random} for details.

Below we consider the RBF kernel on distributions, 
\begin{align*}
K_\bw(P_i,P_j)=\exp\left(-\frac{\norm{p_i-p_j}^2}{2\bw^2}\right),
\end{align*}
where $p_i,p_j$ are the respective densities and $\norm{p_i-p_j}$ is the $L_2$ norm on functions.
We will take the class of mappings we regress to be:
\begin{align}
Y_i = \sum_{j=1}^N \theta_i K_\bw(G_j,P_i) + \epsilon_i \label{eq:2be_model},
\end{align}
where $\norm{\theta}_1<\infty$, $G_j\in \calI$'s are unknown distributions and $\epsilon_i$ is a noise term \cite{oliva2014fast}. Note that this model is analogous to a linear smoother on  some unknown infinite dataset, and is nonparametric. We show that \eqref{eq:2be_model} can be approximated with the 2BE below.

\noindent\textbf{Double-Basis Estimator.}
First note that:
\begin{align*}
\idot{\tp_i}{\tp_j}  =&  \Idot{\sum_{\alpha\in M} a_{\alpha}(\hP_i) \varphi_{\alpha}}{\sum_{\alpha\in M} a_{\alpha}(\hP_j) \varphi_{\alpha}} \\
 =&  \sum_{\alpha\in M} \sum_{\beta\in M}a_{\alpha}(\hP_i) a_{\beta}(\hP_j)\Idot{\varphi_{\alpha}}{\varphi_{\beta}} \\
 =&  \sum_{\alpha\in M} a_{\alpha}(\hP_i) a_{\alpha}(\hP_j) 
 = \Idot{\avec(\hP_i)}{\avec(\hP_j)},
\end{align*}
where $\avec(\hP_i) = (a_{\alpha_1},\ldots,a_{\alpha_s})$, $M = \{\alpha_1,\ldots,\alpha_s\}$, and the last inner product is the vector dot product. Thus,
\begin{align*}
\Norm{\tp_i-\tp_j}_2 = \Norm{\avec_t(\hP_i)-\avec_t(\hP_j)}_2,
\end{align*}
where the norm on the LHS is the $L_2$ norm and the $\ell_2$ on the RHS. 

Consider a fixed $\bw$. Let $\omega_i \stackrel{iid}{\sim} \calN(0,\bw^{-2}I_s)$, $b_i \stackrel{iid}{\sim} \Unif[0,2\pi]$, be fixed. Then,
\begin{align}
\sum_{i=1}^\infty \theta_i K_\bw(G_i, P_0) \approx& \sum_{i=1}^\infty \theta_i K_\bw(\avec(G_i), \avec(P_0)) \nonumber \\
\approx& \sum_{i=1}^\infty \theta_iz(\avec(G_i))^Tz(\avec(\hP_0)) \nonumber \\
=& \left(\sum_{i=1}^\infty \theta_iz(\avec(G_i))\right)^Tz(\avec_t(\hP_0)) \nonumber\\
=& \psi^T z(\avec(\hP_0)) \label{eq:lin_est_approx}
\end{align}
where $\psi = \sum_{i=1}^\infty \theta_iz(\avec(G_i)) \in \R^D$. Thus, we consider estimators of the form
\eqref{eq:lin_est_approx}. I.e. we use a linear estimator in the non-linear space induced by $z(\avec(\cdot))$. In particular, we consider the OLS estimator using the data-set $\{(z(\avec(\hP_i)),Y_i)\}_{i=1}^N$ :
\begin{align}
\hat{f}(\tP_0) \equiv& \hpsi^T z(\avec(\hP_0))\ \where \label{eq:OLSest} \\
\hpsi \equiv& \argmin_\beta \norm{\vY-\bZ\beta}_2^2 
= (\bZ^T\bZ)^{-1}\bZ^T\vY
\end{align}
for $\vY=(Y_1,\ldots,Y_N)^T$, and with $\bZ$ being the $N\times D$ matrix: $\bZ=[z(\avec_t(\hP_1))\cdots z(\avec_t(\hP_N)) ]^T$. 

A straightforward extension to \eqref{eq:OLSest} is to use a ridge regression estimate on features $z(\avec(\cdot))$ rather than a OLS estimate. That is, for $\lambda\geq 0$ let
\begin{align}
\hpsi^T_\lambda \equiv& \argmin_\beta \norm{\vY-\bZ\beta}_2^2 + \lambda \norm{\beta}_2 \label{eq:ridgeest}\\
=& (\bZ^T\bZ+\lambda I)^{-1}\bZ^T\vY.
\end{align}

\subsubsection{Algorithm}
We summarize the basic steps for training the 2BE in practice given a data-set of empirical functional observations $\calD=\{(\calX_i,Y_i)\}_{i=1}^N$, parameters $\bw$ and $D$ (which may be cross-validated), and an orthonormal basis $\{\varphi_i\}_{i\in\Z}$ for $L_2([a,b])$.
\begin{enumerate}
  \item Determine the sets of basis functions M for approximating $p$. This may be done via cross validation of density estimates (see \cite{oliva2014fast} for more details).
  \item Let $s=|M|$, draw $\omega_i \stackrel{iid}{\sim} \calN(0,\bw^{-2}I_s)$, $b_i \stackrel{iid}{\sim} \Unif[0,2\pi]$ for $i\in\{1,\ldots,D\}$; keep the set $\{(\omega_i,b_i)\}_{i=1}^D$ fixed henceforth.
  \item Let $\{\alpha_1,\ldots,\alpha_s\}=M$. Generate the data-set of random kitchen sink features, output projection coefficient vector, response pairs $\{(z(\avec(\hP_i)),Y_i)\}_{i=1}^N$. Let $\hat{\psi} = (\bZ^T\bZ+\lambda I)^{-1}\bZ^T\vY \in\R^{D} $ where $\bZ=[z(\avec(\hP_1))\cdots z(\avec(\hP_N)) ]^T \in\R^{N \times D}$, and $\lambda$ may be chosen via cross validation. Note that $\bZ^T \vY$ and $\bZ^T \bZ$ can be computed efficiently using parallelism.
  \item For all future query input functional observations $\hP_0$, estimate the corresponding response as $\hat{f}(p_0) = \hat{\psi}^T z(\avec(\hP_0))$.
\end{enumerate}

\subsection{2BE for Redshift Prediction}
We divide simulation snapshots into 16 \mpc length sub-cubes, for a total of 512 sub-cubes per simulation snapshot. Each sub-cube is then rescaled to be the unit box. We treat each sub-cube as a sample $\calX_i$ with a response  $Y_i$, of the redshift it was observed at. In total, a training set of approximately 600K (sample $\calX_i$, response $Y_i$) pairs was used for constructing our model. A total of 130 simulation snapshots were held out. Test accuracies were assessed by averaging the predicted response in the boxes of each held-out snapshot. 

We used 20K random features, D, as in Eq.~\eqref{eq:rksrfeat}. We used the cosine basis, \ie the tensor product in Eq.~\eqref{eq:basisprod} of: $\varphi_0(x)=1$, and $\varphi_k(x)=\sqrt{2}\cos(k\pi x)$ for $k\geq 1$. The set of basis functions, $M$ \eqref{eq:coef-hat}, was taken to be $M = \{\alpha \in \mathbb{N}^3\ :\ \norm{\alpha} \leq 18 \}$ via rule of thumb. The free parameters $\bw$, the bandwidth, and $\lambda$, the regularizer, were chosen by validation on a held-out portion of the training set. In total the 2BE model's parameters $\psi$, totaled 20K dimensions. 

\section*{Future Directions} 
We demonstrated that machine learning techniques can produce accurate estimates of the cosmological parameters from simulated dark matter distributions, which are highly competitive with standard analysis techniques.
In particular the advantage of conv-nets on small-scale boxes shows that convolutional features that carry higher order correlation information provide high fidelity and 
could produce low variance estimates of the cosmological parameters.

The eventual goal is to use such models to estimate the parameters of our own Universe, where we only have access to the distribution of ``visible'' matter.  This introduces extra complexities as galaxies and clusters are biased tracers of the underlying matter distribution.  Furthermore, the direct simulation of galaxy clusters are highly complex.
In the next step, we would like to evaluate and establish the robustness of these models to variations \textit{across} simulation settings, before applying proper models to Sloan Digital Sky Survey data~\cite{alam2015} that observes the distribution of galaxies at large scales.


\section*{Acknowledgements}
We would like to thank Hy Trac for providing the simulations for the second set of experiments.
We also like to thank the anonymous reviewers for their helpful feedback.
The research of SR was supported by the department of energy grant DE-SC0011114.

{
\bibliography{refs.bib}
\bibliographystyle{icml2016}
}
\clearpage
\appendix
\section{Spatial Scale of the Cubes}\label{sec:scales}
\begin{figure}[h!]
\centering
\hbox{
\subfigure[sub-cubic volume of $64^3  \mpc$]{\label{fig:pred_512}\includegraphics[width=.4\textwidth]{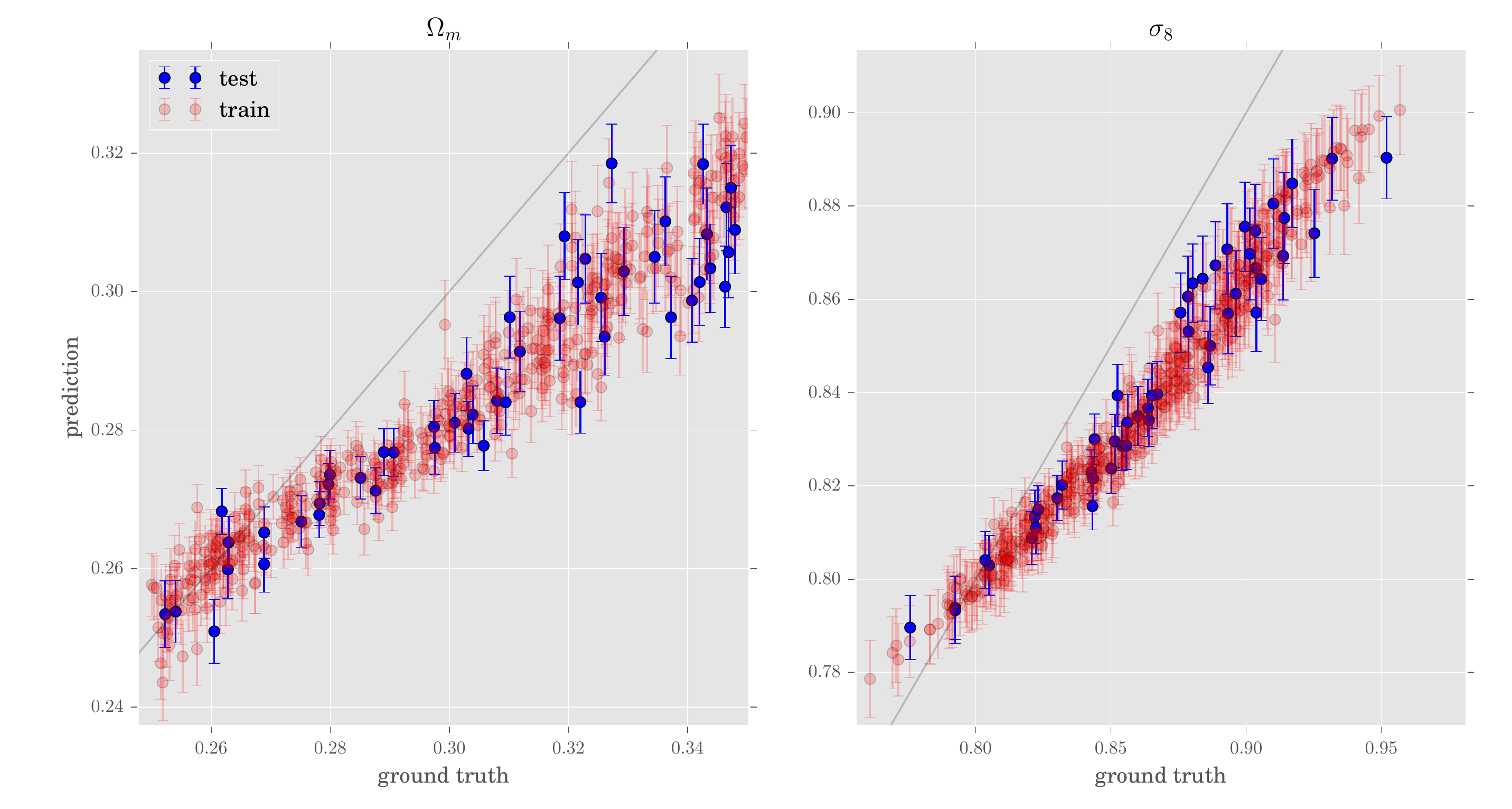} }}
\hbox{
\subfigure[sub-cubic volume of $128^3  \mpc$]{\label{fig:pred_std}\includegraphics[width=.4\textwidth]{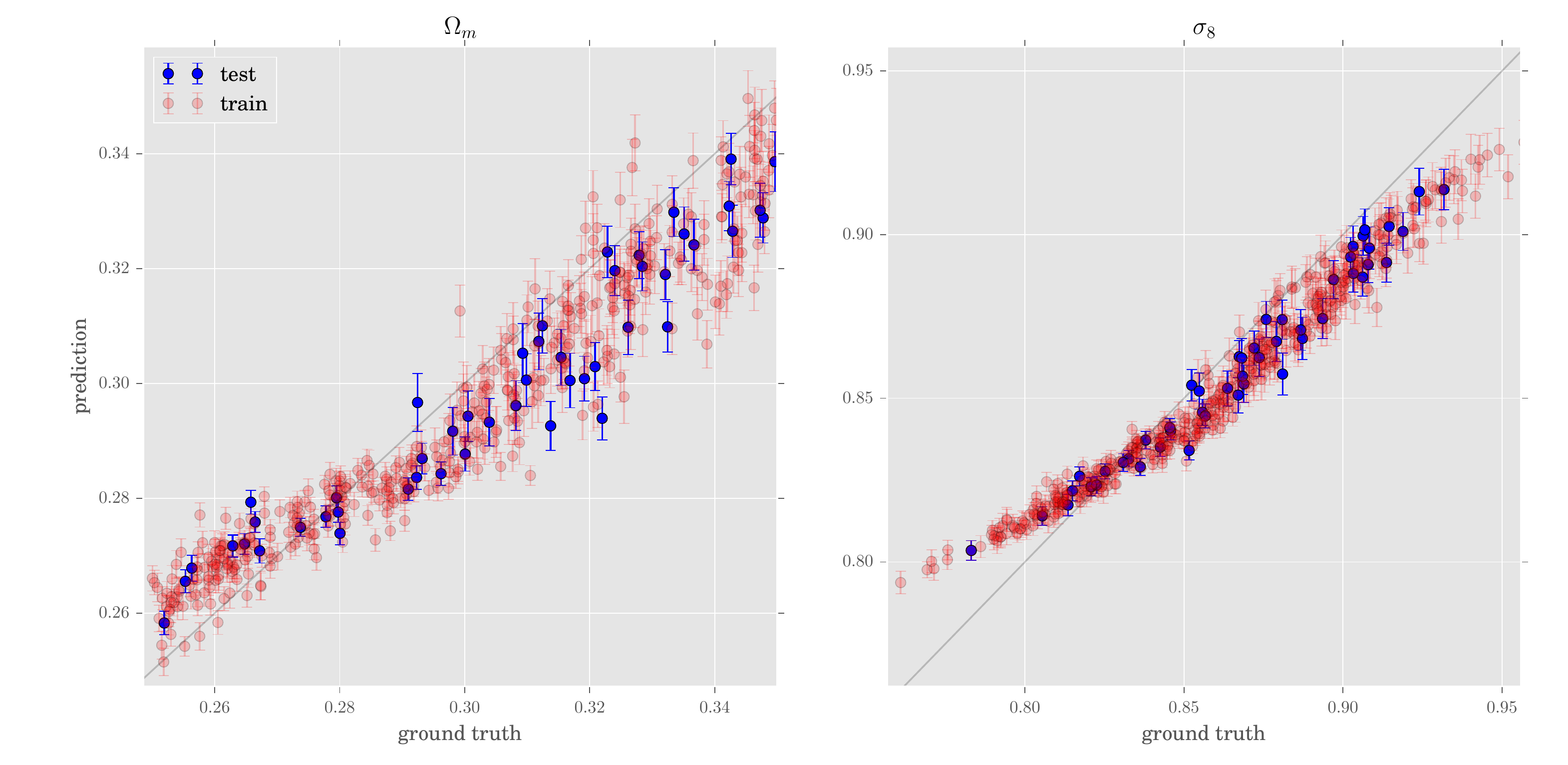} }}
\hbox{
\subfigure[sub-cubic volume of $(256^3 \mpc$]{\label{fig:pred_128}\includegraphics[width=.4\textwidth]{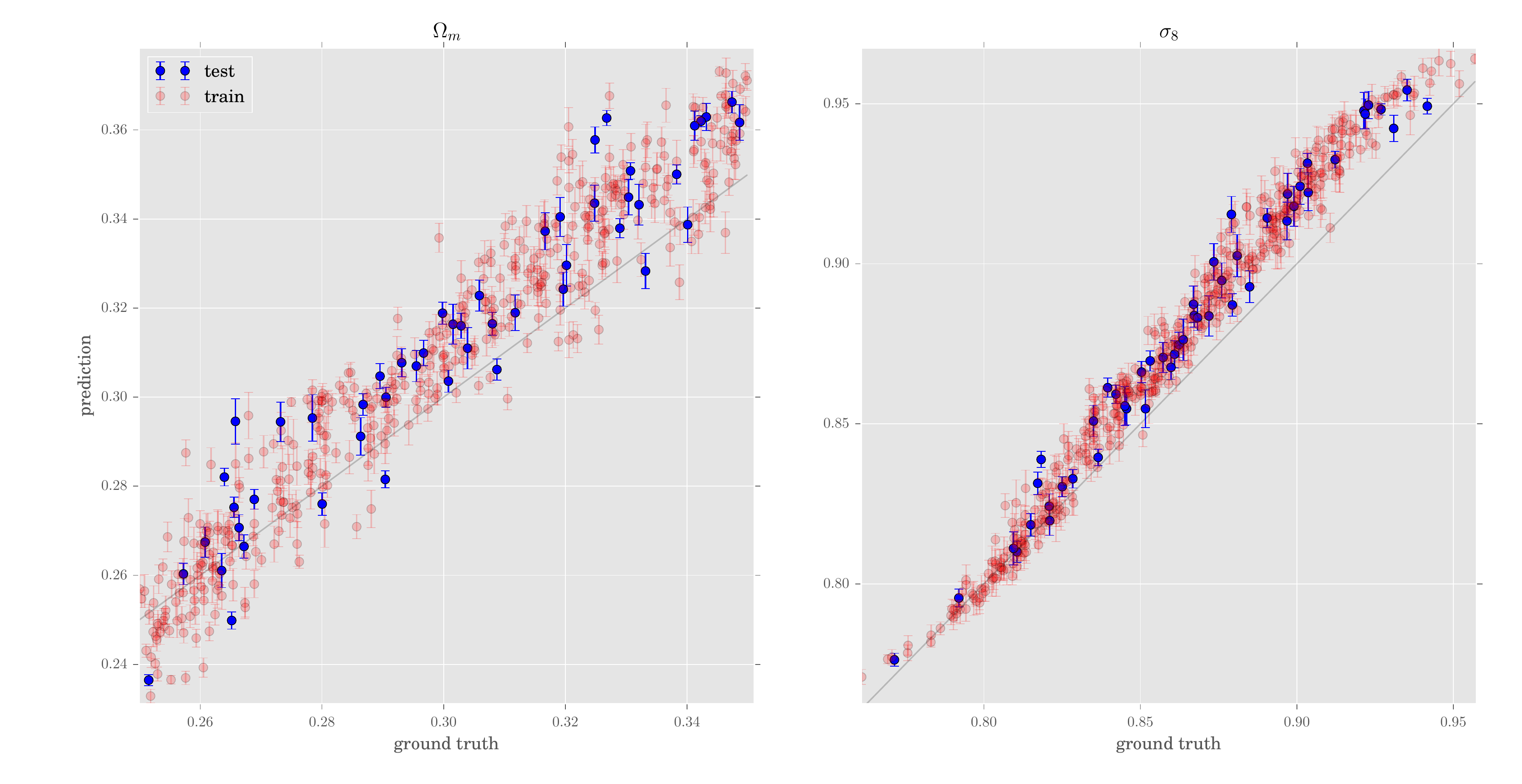} }}
\caption{\small{
Prediction and ground truth using a) small; b) medium and ;c) large sub-cubes.
The error-bar shows the standard deviation over predictions made by sibling sub-cubes. 
}
}\label{fig:preds_all}
\end{figure}

However, increasing the size of cubes comes at the cost of less training/test instances.
We evaluated the effect of scale by using different quantizations of the original $512^3 (\mpc)^3$ cubes. 
The results of \cref{sec:results} use a 3D histogram with $256^3$ voxels divided into 
$64^3$-voxel sub-cubes. We also tried 3D histograms with $512^3$ and $128^3$ voxels, with similar $64^3$-voxel sub-cubes. We then used the same conv-net for training. This resulted
in using $2^3=8$ times less or more instances. 
 \Cref{fig:preds_all} compares the prediction accuracy under different spatial scales.
Error-bars show one standard deviation for the predictions made using 
different sub-cubes that belong to the same cube (sibling sub-cubes).  
Interestingly, these predictions for sibling sub-cubes are also consistent, having a small standard deviation for both parameters ($\Omega_m$ and $\sigma_8$).

Moreover, change of the spatial volume of sub-cubes does not seem to significantly affect the prediction accuracy.  We are able to make predictions with similar accuracy using sub-cubes with both smaller and larger spatial scales.

\end{document}